\documentclass[12pt]{article}
\usepackage{hyperref}

\usepackage{graphicx}

\begin{document}

\title{\begin{flushright} {\footnotesize IFUP-7/03} \end{flushright} Reaction Mechanisms
with Exotic Nuclei\footnote {Invited talk given at the Symposium on Nuclear Clusters,
Rauischholzhausen,  Germany, 5-9 August 2002. }}
 
\author{Angela Bonaccorso \footnote{In collaboration with G. Blanchon, D.M.
Brink, J. Margueron, N. Vinh Mau. }
 \\Istituto Nazionale di Fisica Nucleare, Sez. di Pisa, \\ 
56127 Pisa, Italy\\[0.2ex] }

 \maketitle

\abstract{This talk examines a number of reaction  mechanisms for scattering initiated by an exotic
projectile. Comparisons are made with recent experimental data, in order to extract
information on the peculiarity of the nuclear structure under extreme conditions and to
test the accuracy of the available
theoretical methods. Predictions for future experiments are also made. }

\maketitle

\section{Introduction}\label{intro}
Nuclei far from  the stability valley are often called "exotic" because they exhibit 
properties rather different from those of nuclei in the rest of the nuclear chart
\cite{eric}. Most of them are neutron rich and unstable  against $\beta$-decay. It is
interesting to study them because they give information on the structure of matter under
extreme conditions and allow to test nuclear models otherwise based only on
properties of stable nuclei.

 From the theoretical point of view the most
interesting characteristic of medium-light unstable nuclei is the fact that  single
particle degrees of freedom dominate both in the structure description as well as in
the reaction studies.  So far the most
studied cases have been those of light nuclei like $^{6}He$, $^{11}Be$ $^{11}Li$
which exhibit the so called halo. $^{19}C$ is another interesting candidate still
under investigation \cite{va,nak,typ}.  Due to the fact
that the last neutron or couple of neutrons (in
$^{6}He$ and
$^{11}Li$ ), are weakly bound, with a separation energy of around 0.5MeV, the wave functions of such  valence neutrons
exhibit long tails which extend well outside the nuclear potential well and  a
large part of the single particle strength is already  in the continuum. From the structure point of view the
dominant feature is the appearance of intruder states in the single particle
level scheme and the strong coupling between deformed cores and valence
particles. 
New techniques are needed to study these nuclei, which combine and unify the
traditional treatment of bound and continuum scattering states. In this respective
reaction theories  like the transfer to the continuum 
\cite{bb,bb1,bb4}  can be very useful. 

 Therefore, as in the early stages of Nuclear
Physics, research on light exotic nuclei has concentrated on studying
elastic scattering \cite{bc}-\cite{so} and spectroscopic properties like the determination of
single particle state energies,  angular momenta and spectroscopic factors
\cite{ta,be12,je}.

\section{Reaction models for structure studies}

There are several cross sections that are measured and calculated in
   the models. Elastic scattering angular distributions have 
been measured. The comparison between the scattering of a halo nucleus of mass A with that
of the nucleus (A-1) has shown a considerable depletion of cross section which has been
explained as due to the breakup channel\cite{bc}. 
On the other hand early measurements concentrated on total reaction cross sections 
for the extraction of nuclear
radii \cite{tan,kob} and were also used to disentangle  the single particle level
sequence of halo nuclei \cite{menic}. 

The next
simplest measurement is the single-neutron removal
   cross section, in which only the projectile residue, namely the
   core with one less nucleon, is observed in the final state. This information
together with the calculated cross sections \cite{va,nak},\cite{ta,be12,je} has been
used to extract single particle spectroscopic factors as in traditional transfer
reactions. Besides
   the integrated removal cross section, denoted by $\sigma_{-n}$, the
   differential momentum distribution $d^3\sigma/d k^3$ is also
   measured. A particularly useful cross section is $d \sigma /d
   k_z$, the removal cross section differential in longitudinal momentum. It has been
used to determine the angular momentum and spin of the neutron initial state \cite{je} in
a way similar to that proposed in \cite{bb4,tiina}.
   If the final state neutron can also be measured, the corresponding
   coincident cross section $A_p \rightarrow (A_p-1) +n$ is called the
   diffractive (or elastic ) breakup cross section if the interaction responsible for the
removal is the neutron-target nuclear potential \cite{anne,ab}. In the case of heavy
targets the coincident cross section contains also the contribution from Coulomb breakup
due to the core-target Coulomb potential which acts as an effective force on the
neutron. This observable is very useful to disentangle the reaction mechanism
\cite{jer}. The difference between the removal and  coincident cross sections is called
the stripping (or absorption) cross section.

All theoretical methods used so far rely on a basic approximation to describe the collision with
   only the three-body variables of nucleon coordinate, projectile
   coordinate, and target coordinate. Thus the dynamics is controlled by the
   three potentials describing nucleon-core, nucleon-target, and core-target
   interactions. In most cases the projectile-target relative motion is treated
   semiclassically by using a trajectory of the center of the projectile
   relative to the center of the target ${\bf R}(t)={\bf b_c}+{\bf v}t$
   with constant velocity $v$ in the $z$ direction and impact parameter
   {\bf b$_c$} in the $xy$ plane. 

\subsection{Nuclear-Coulomb elastic breakup. }\label{na}

A full description of the treatment of the scattering
equation for a projectile which decays by single neutron breakup following its
interaction with the target, can be found in \cite{bb1,jer}. There it was shown that
within the semiclassical approach for the projectile-target relative motion, the
amplitude for a
   transition from a  nucleon bound state $\psi_i$ in the projectile to a final
continuum state
$\psi_f$ is given by 
   \begin{equation}A_{fi}={1\over i\hbar}
   \int_{-\infty}^{\infty}dt<\psi_{f} (t)|V({\bf r})|\psi_{i}(t)>,\label{1}\end{equation}
where $V$ is the
    interaction responsible for the neutron transition to the continuum. 

For light targets the recoil effect due to the projectile-target Coulomb potential can be
neglected and the interaction responsible for the reaction is mainly the neutron-target
nuclear potential. In the case of heavy targets the dominant reaction mechanism is
Coulomb breakup.  The Coulomb force does not act directly on the neutron but it affects
it only indirectly by causing the recoil of the charged core.  Therefore the neutron is
subject to an effective force which gives rise to an effective dipole Coulomb potential
$V_{eff}({\bf r,R}(t))$. 
 In ref.\cite{jer} it was shown that the combined effect of the nuclear and Coulomb
interactions to all orders can be taken into account by using the potential
${V}=V_{nt}+V_{eff}$ sum of the neutron-target optical potential and the Coulomb dipole
potential.   If for the neutron final continuum  wave function we take a distorted
wave of the  eikonal-type, then   the amplitude becomes :
\begin{equation}
A_{fi}\left(  \mathbf{k,}\mathbf{b_c}\right)  =\frac{1}{i\hbar}\int
d^{3} {\bf r} \int dte^{-i{\bf k \cdot r}+i\omega t} e^{\left(  \frac{1}{i\hbar
}\int_{t}^{\infty}{V}\left(  {\bf r},t^\prime\right)  dt^\prime\right)  }
{V}\left(  {\bf r},t\right)  \phi_{l_im_i}\left(  \mathbf{r}\right)
\label{amp1}%
\end{equation}
where $\omega=\left(  {{\varepsilon_f}^{\prime}}-\varepsilon_{0}\right)  /\hbar
$ and $\varepsilon_0$ is the neutron initial bound state energy  while
${{\varepsilon_f}^{\prime}}$ is the neutron-core final continuum energy. 
Eq.(\ref{amp1}) is appropriate to calculate the coincidence cross sections $A_p
\rightarrow (A_p-1) +n$ discussed in the previous section. Finally the
differential probability with respect to the neutron energy and angles can be
written as
${d^3P_{nc}(b_c)\over d{{\varepsilon_f}^{\prime}}\sin\theta d\theta d\phi}
={1\over 8\pi^3}{m k_n\over \hbar^2}{1\over
2l_i+1}\Sigma_{m_i}| A_{fi}|^2 .\label{anc}$
where $A_{fi}$ is given by Eq.(\ref{amp1}) and we have averaged over the neutron
initial state.

   The effects associated with the core-target interaction will be included by 
   multiplying the above probability by $P_{ct}(b_c)=|S_{ct}|^2$ \cite{bb4} the
probability for the core to be left in its ground state, defined in
terms of a core-target S-matrix function of  $b_c$, the core-target distance of
closest approach. A simple parameterization is
$P_{ct}(b_c)=e^{(-\ln 2 exp[(R_s-b_c)/a])}$, where the strong absorption radius $R_s\approx 1.4
 (A_p^{1/3}+A_t^{1/3}) fm$ is defined as the distance of closest approach for a trajectory that is 50\%
absorbed from the elastic channel and
$a=0.6fm$ is a diffusness parameter.

  Thus the double differential cross section  is 
    \begin{equation}
     {d^2\sigma\over  {d{{\varepsilon_f}^{\prime}} d\Omega}}=C^2S
    \int_0^{\infty} d{\bf b_c} {d^2 P_{nc}({\bf k},b_c)\over
d{{\varepsilon_f}^{\prime}}d\Omega} 
    P_{ct}(b_c), \label{cross} \end{equation}
    (see Eq. (2.3) of \cite{bb4}) and $C^2S$ is the spectroscopic
   factor for the initial single particle orbital.

\subsection{Nuclear elastic and absorptive breakup. }\label{nab}

Inclusive cross sections in which only the core
with $(A_p-1)$ nucleons is detected need to take into account also the
absorption of the neutron by the imaginary part of the n-target optical
potential. For
such reactions the Coulomb recoil effect can be neglected but the distorted eikonal-type wave function used in
Eq.(\ref{amp1}) is not accurate enough, in particular if the final continuum states are single particle
resonances in the target plus one neutron nucleus. Then
 a distorted final neutron wave function, calculated by an optical model  
will be used. Also since the neutron is not detected one integrates over the neutron angles.  Thus, according
to \cite{bb1} the final neutron probability energy spectrum with respect to the target reads
\begin{equation}{dP\over d\varepsilon_f} \approx {1\over
2}\Sigma_{j_f}(|1-\langle S_{j_f}\rangle |^2+1-|\langle S_{j_f}\rangle |^2)
(2j_f+1)(1+F_{l_f,l_i,j_f,j_i})B_{l_f,l_i}. \label{dpde}\end{equation}

\begin{equation}B_{l_f,l_i}=\left [{1\over mv^2}\right ]{1\over
k_f}|C_i|^2 {e^{-2\eta b_c}\over 2\eta b_c}M_{l_fl_i},\label{B}\end{equation}
where $S_{j_f}$ is the neutron-target optical model S-matrix,
$F_{l_f,l_i,j_f,j_i}$ is an $l$ to $j$ recoupling factor, $\eta$ is the
transverse component of the neutron momentum which is conserved in the
neutron transition, $b_c$ is the core-target impact parameter,
$C_i$ is the initial state asymptotic normalization constant and $M_{l_fl_i}$
is a factor depending on the angular parts of the initial and final  wave
functions, $v$ is the relative motion velocity at the distance of closest approach.
\section{Applications}

We are going to discuss now a series of calculations aimed at extracting spectroscopic 
information on one-neutron and two-neutron halo nuclei.

\subsection{ Neutron  energy distributions in Coulomb breakup. }

\begin{figure}[htb]
\hskip .5in\includegraphics [scale=0.5,angle=-90]{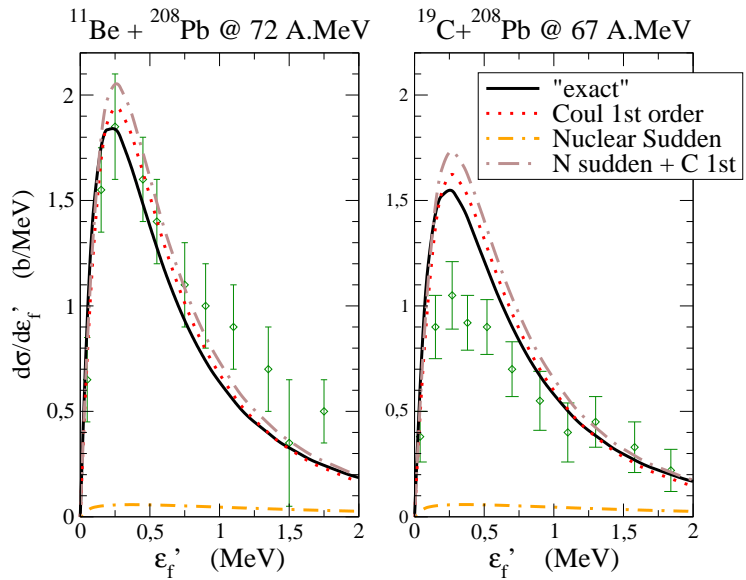}
   \caption{Neutron-core final energy distribution after nuclear-Coulomb breakup.}\label{eps2}
\end{figure}

$^{11}Be$ is probably the best known one-neutron halo nucleus since experimental information has
been available for long time \cite{mill}. The ground state is a $2s_{1/2}$ state with
separation energy of 0.5MeV and spectroscopic factor $C^2S=0.77$. Therefore it has been
used as a test case for reaction models which use the above basic structure information as
input. A very comprehensive study  of one neutron breakup mechanism, cross section and momentum distributions 
can be found in \cite{ta,anne,ab,abb,abfc}.

On the other hand a more recent work \cite{jer} has improved  the previous knowledge 
of the breakup reaction, by
studying the Coulomb-nuclear interference effects according to Eq.(\ref{amp1}).
We would like to report here on new calculations that we have recently performed by 
using Eq.(\ref{amp1}) to study higher order effects. We have tested
three limits of Eq.(\ref{amp1}). The first is the sudden
approximation in which  $\omega=0$ and Eq.(\ref{amp1}) can be calculated with  nuclear and Coulomb to all
orders. We call the corresponding amplitude $A_{sudd}^{all-ord}$. Then we have studied the first order
approximation for the Coulomb term in which
$e^{\left( 
\frac{1}{i\hbar }\int_{t}^{\infty}{V_{eff}}\left(  {\bf r},t^\prime\right)  dt^\prime\right) 
}=1$ but the
$\omega t$ term is kept (this is the standard first order perturbation theory
amplitude $A^{pert}$) and  finally the sudden approximation restricted to first order  giving
$A^{pert}_{sudd}$. The main results of our new calculations are shown in Fig.(1a) and (1b)
which give the neutron final energy spectrum with respect to the core  for breakup of
$^{11}Be$ and
$^{19}C$ on $^{208}Pb$ at 72 A.MeV and 67 A.MeV respectively. Experimental data are from
\cite{nak}. Preliminary calculations indicate that for $R_s < b_c < 50fm $ the results
obtained with $A^{pert}$ are equal to those obtained with $A^{pert}_{sudd}$, thus
showing that at small impact parameters the sudden approximation is valid but higher order terms need to be
considered. On the other hand for $b_c >50fm $ we find that higher order effects are negligible since using
$A_{sudd}^{all-ord}$ or $A^{pert}_{sudd}$ does not give any difference. Then we can
conclude that at large $b_c$ perturbation theory is valid. Thus in the figures we give
by the dotted curves  the results of the simple first order perturbation theory  while the
solid curves are the all order calculations according to an amplitude defined as
$A^{'exact'}=A_{sudd}^{all-ord}+A^{pert}-A^{pert}_{sudd}$, valid at all core-target
impact parameters and  not giving rise to divergences in the final integral over
impact parameters in Eq.(\ref{cross}).
 In the case of
$^{11}Be$ the theoretical calculations have been multiplied by the known spectroscopic
factor, while for $^{19}C$ we have used unity spectroscopic factor and a neutron
separation energy for the 2s state of 0.5MeV. As expected, and already shown by other
authors the effects of higher order terms are to reduce the peak cross section. Analysis
of the type presented in this section have been used to extract spectroscopic factors.
From our calculations we would extract $C^2S=0.70$ for the 2s-state of $^{11}Be$ and
$C^2S=0.65$ for $^{19}C$ assuming   in both cases  a separation energy of 0.5MeV.

\subsection{$^{10}Li$ spectrum and $^{11}Li$ properties.}
\begin{figure}[htb]
\vspace*{-12pt}
 \includegraphics[scale=0.5,angle=90]{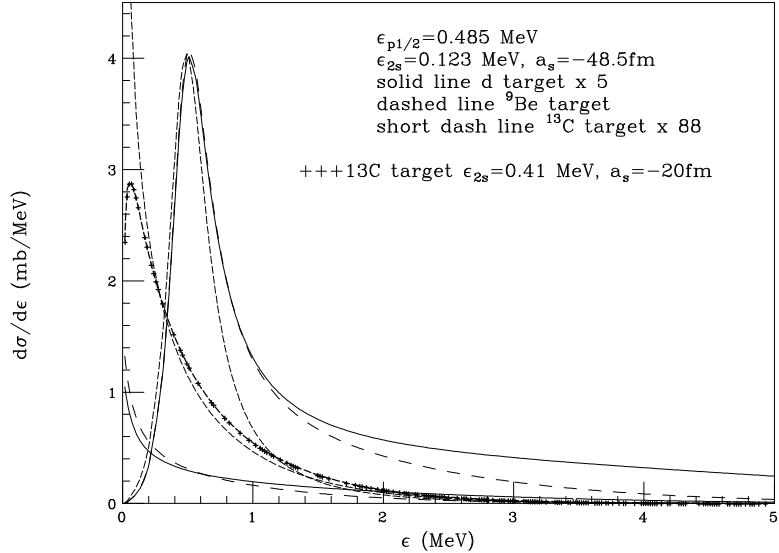}
\vspace*{-12pt}
   \caption[]{Neutron-$^{9}$Li relative energy spectra for transfer to the s and
p continuum states in $^{10}$Li.}\label{eps4}
\end{figure}

We discuss now the results of a possible reaction aiming at clarifying the structure $^{11}Li$ which 
  has been a challenge for long time \cite{menic,tho}-\cite{sf}. This nucleus and
$^{6}He$ are two-neutron halo nuclei. They are special because their corresponding A-1
systems are unbound and it is thanks to the pairing force acting between the two neutrons
that they become bound. $^{11}Li$ has been very difficult to study from the experimental
point of view  because the ground state of $^{10}Li$ is unbound
\cite{cha}-\cite{sf}, and
  one of the available states for the valence
neutron is  a $2s_{1/2}$ virtual state which does not even have a centrifugal barrier.
$^{6}He$ is different in this respect because the $p_{3/2}$ ground state resonance of $^{5}He$ has a
width of about 600KeV corresponding to a lifetime of about 300fm/c \cite{6he} and its decay in flight has been
clearly observed. Recently a pickup experiment $d(^{11}Be,^{3}He)^{10}Li$ 
\cite{sf} has definitely confirmed the earlier hypothesis that the ground state of
$^{10}Li$ is the 2s virtual state and that the $1p_{1/2}$ orbit gives an excited state. 
Three body models of $^{11}Li$ need as a fundamental ingredient the n-core
(n-$^{9}$Li) interaction, which in turn determines the energies of the low energy unbound
states in $^{10}$Li. Following ref.\cite{pvm} the two neutron hamiltonian is
$H_{2n}=h_{1}+h_{2}+V_{nn}$.
$V_{nn}$ is the zero-range paring interaction. The single
neutron hamiltonian is 
$h=t+ V_{cn}$ where $t$ is the kinetic energy and $V_{cn} =V_{WS}+\delta V$ is the
neutron-core interaction. It is given by the usual Woods-Saxon potential plus
spin-orbit plus a correction
$\delta V$ which originates from particle-vibration couplings.  They are important for low
energy states but can be neglected at higher energies. If Bohr and Mottelson collective
model is used for the transition amplitudes between zero and one phonon states, then 
 $
\delta V(r)=16\alpha {e^{{{2(r-R)}/ a}}/({1+e^{{{(r-R)}/
 a}}})^4 }$\
where $R\approx r_0A^{1/3}$. According to \cite{pvm}  the best parameters
for the n-$^{9}$Li interaction  given in Table 1.
\begin{table}[t]
\caption{Woods-Saxon and spin-orbit potential parameters}
\footnotesize
\begin{center}
\footnotesize
\begin{tabular}{ccccc}
\hline
\hline
$V_0$      & $r_0$   & $a$    & $V_{so}$ & $a_{so}$ \\
 ($MeV$)   &($fm$) & ($fm$) & ($MeV$)  & ($fm$)   \\
\hline
&&&&\\
39.83      &1.27   & 0.75   &  7.07    & 0.75     \\
\ \        & \ \   & \ \    & \ \      & \ \      \\
\hline
\hline
\end{tabular}
\end{center}
\end{table}
The corresponding energies obtained for the 2s and 1p$_{1/2}$ state are given in Table
2, together with the values of the strength $\alpha$ of the correction potential $\delta V$.

\begin{table}[t]
\caption{Energies of the s and p states, width of unbound p-state, scattering length
of the s-state and strength parameter for the $\delta V$ potential. (a) bound-state like
calculation, (b) scattering state calculation }\vskip .2in
\footnotesize
\begin{center}
\footnotesize
\begin{tabular}{lccccc}
\hline
\hline\
                        & (a) & (b)&$\Gamma$(MeV)&$a_s$(fm)& $\alpha(MeV)$\\
$\epsilon_{2s}(MeV)$       & 0.123 & 0.17 &&-48.5&-13.3\\
&&0.45&&-20&-14.0\\
$\epsilon_{1p_{1/2}}(MeV)$ & 0.485 & 0.595 &0.48&&3.3\\
\hline
\hline
\end{tabular}
\end{center}
\end{table}
 It would be therefore extremely interesting and important if an experiment could
determine the energies of the two unbound $^{10}$Li states such that the interaction
parameter could be deduced. Two $^{9}Li(d,p)^{10}Li$ experiments have recently been
performed. One at MSU at 20 A.MeV \cite{santi} and the other at the CERN REX-ISOLDE facility
at 2 A.MeV\cite{bj}. For such transfer to the continuum reactions the theory underlined in
Section (2.2) is very accurate. We present in the following our predictions. In
order to study the sensitivity of the results on the target and on the energies assumed
for the s and p states, we have calculated the reaction $^{9}Li(X,X-1)^{10}Li$ at 2
A.MeV for three targets d, $^{9}$Be, $^{13}$C. 
 The  $^{13}$C target has been chosen
because in such a case the neutron transfer to the 2s state in $^{9}$Li would be a
spin-flip transition which as it is well known  are enhanced at low incident
energy. For the other two cases the transfer to the 2s state is a non spin-flip
transition which is hindered. We show in Fig.2 the neutron energy spectrum relative to 
$^{9}$Li obtained with the interaction and single particle energies of Tables 1 and 2. In the
case of the 2s virtual state we give also the scattering length obtained as
$a_s=-  \mathrel{\mathop{lim}\limits_{k \to
0\hspace{.28em}}}{tan\delta_0\over k} $. We define  the resonance energy of the
p-state and the energy of the virtual s-state as the energy at which
$\delta_l=\pi/2$ and therefore $Im S_l$ changes sign. We have checked that in this way
we get the s continuum state at the same positive energy if we solve the Sch\"oedinger
equation in a box  or if we solve it with the
proper scattering boundary conditions. This is also the energy at which
$|1-\langle S_{j_f}\rangle|^2$ in eq.(\ref{dpde}) gets its maximum value. It is important to
stress such a definition in the case of the s-virtual state. This is because our
prescription gives different energies than those obtained using the relation
$\varepsilon_{2s}={\hbar^2\over 2 m a_s^2}$.  The results of Fig.2 show that the
peak of the p-state will determine without ambiguity the position of the
p-state in a target independent way. The width instead is modified by the
reaction mechanism, but it can however be obtained from the theory (actually
from the phase shift behavior) once that the energy is fixed.  For the s-state
we see that there is a larger probability of population in the spin-flip
reaction initiated by the carbon target. A measure of the line-shape (or
spectral function) and absolute value of the cross section will determine the
energy of the state also in this case. The integral over energy of the energy distribution will determine
the spectroscopic factor of the state. In this case there is no spreading of the single particle state since
the n-$^9$Li interaction is real at such low energies. In fact the first excited state of $^9$Li is at
$E^*=2.7MeV$. In order to demonstrate the sensitivity of the model calculation to the energy of the state we
show in Fig.2 by the crosses the result obtained if the s-state is located at
$\varepsilon_{2s}=0.45 MeV$ corresponding to $a_s=-20 fm$. In this case a clear peak
appears even if located at very small energy. The fact that a peak appears or does not
appear in the transfer spectrum, depends on the relative behavior of the two terms
$|1-\langle S\rangle|^2$ and B($j_i,j_f$..) in Eq.(4). The $|1-\langle S\rangle|^2$ term has always a maximum
value equal to 4 at the energy of the state, while the B-term has a divergent-like
behavior as the energy approaches zero. Therefore we can conclude that if a transfer to
the continuum  experiment could measure with sufficient accuracy (energy resolution)
the line-shapes or energy distribution functions for the s and p-states in $^{10}$Li our
theory would be able to fix unambiguously the energies of the states. Those in turn
could be used to test microscopic models of the n-$^{9}$Li interaction.

\section{Conclusions and future challenges}

It is clear from what we have discussed in this paper that physics with radioactive beams
is an extremely fascinating field in which the interplay between the understanding of the
nuclear structure and that of the reaction mechanism is very strong and  an enormous
number of progress has been made in the last few years. There are however a number of
improvements both experimental as well as theoretical that need to be pursued. Almost
all experiments so far performed have been inclusive with respect to the target. 
Up to date few experiments with full kinematics reconstruction have been performed such as
those of Galin and collaborators
\cite{joel}. But targets like
$^{9}Be$ which has been widely used, are themselves very weakly bound and probably undergo breakup
following the interaction with radioactive beams. Data presently available most probably
contain such contributions. The picture contained in Eq.(\ref{cross}) needs therefore to
be modified to take into account more complicated situations in which the core-target
scattering is NOT elastic. Spin coupling effects and final neutron energy dependence have been
neglected in most of the theoretical approaches. The discussion about $^{34}Si$ in \cite{je} clearly
shows that if we are going to study heavier systems in which binding energies might not
be so small to generate halos but rather neutron skins, such effects will need to be
taken into account. On the other hand two-neutron halo breakup of $^{11}Li$ has been
treated as a process in which the two neutrons are emitted simultaneously in a single
breakup process. This is in fact not correct for the second neutron which decays in
flight from a resonant state, as seen for $^{6}He$, and therefore cannot be described by
a breakup form factor of the same type as for the first neutron.

\end{document}